\documentclass[conference]{IEEEtran}
\IEEEoverridecommandlockouts
\usepackage{cite}
\usepackage{amsmath,amssymb,amsfonts}
\usepackage{algorithmic}
\usepackage{graphicx}
\usepackage{textcomp}
\usepackage{xcolor}
\usepackage{float}
\usepackage{subcaption} 
\def\BibTeX{{\rm B\kern-.05em{\sc i\kern-.025em b}\kern-.08em
    T\kern-.1667em\lower.7ex\hbox{E}\kern-.125emX}}

\usepackage{pifont}
\usepackage{subfig}
\usepackage{colortbl}
\usepackage{braket}
\usepackage{comment}

\newcommand{\CellWithForceBreak}[2][c]{
\begin{tabular}[#1]{@{}c@{}}#2\end{tabular}}

\usepackage{balance}
\usepackage{xcolor}
\usepackage[ruled,vlined,noend]{algorithm2e}

\SetCommentSty{mycommfont}
\SetAlFnt{\small}
\SetAlCapFnt{\small}
\SetAlCapNameFnt{\small}

\usepackage{soul}
\def\BibTeX{{\rm B\kern-.05em{\sc i\kern-.025em b}\kern-.08em
    T\kern-.1667em\lower.7ex\hbox{E}\kern-.125emX}}

\begin{document}

\title{TetrisLock: Quantum Circuit Split Compilation \\with Interlocking Patterns}
\author{
    \IEEEauthorblockN{Qian Wang\IEEEauthorrefmark{1}, Jayden John\IEEEauthorrefmark{1}, Ben Dong\IEEEauthorrefmark{1},  Yuntao Liu\IEEEauthorrefmark{2}}
    \IEEEauthorblockA{\IEEEauthorrefmark{1}Department of Electrical Engineering, University of California, Merced, CA, USA}
    \IEEEauthorblockA{\IEEEauthorrefmark{2}Department of Electrical and Computer Engineering, Lehigh University, PA, USA}
    \IEEEauthorblockA{qianwang@ucmerced.edu, jjohn92@ucmerced.edu, cdong12@ucmerced.edu, yule24@lehigh.edu}

}

\maketitle



\begin{abstract}
In quantum computing, quantum circuits are fundamental representations of quantum algorithms, which are compiled into executable functions for quantum solutions. Quantum compilers transform algorithmic quantum circuits into one compatible with target quantum computers, bridging quantum software and hardware. However, untrusted quantum compilers pose significant risks. They can lead to the theft of quantum circuit designs and compromise sensitive intellectual property (IP). 
In this paper, we propose TetrisLock, a split compilation method for quantum circuit obfuscation that uses an interlocking splitting pattern to effectively protect IP with minimal resource overhead. Our approach divides the quantum circuit into two interdependent segments, ensuring that reconstructing the original circuit functionality is possible only by combining both segments and eliminating redundancies. This method makes reverse engineering by an untrusted compiler unrealizable, as the original circuit is never fully shared with any single entity. 
Also, our approach eliminates the need for a trusted compiler to process the inserted random circuit, thereby relaxing the security requirements. Additionally, it defends against colluding attackers with mismatched numbers of qubits, while maintaining low overhead by preserving the original depth of the quantum circuit. We demonstrate our method by using established \textit{RevLib} benchmarks, showing that it achieves a minimal impact on functional accuracy (less than 1\%) while significantly reducing the likelihood of IP inference. 

\end{abstract}

\maketitle

\vspace{-2mm}
\section{Introduction}
\vspace{-2mm}

The advancement of quantum computers ushers a new age of computational power and methodologies.
At the core of each quantum computer are qubits, which differ from classical bits by operating based on principles of quantum mechanics. Unlike classical bits, which are either 0 or 1, qubits can exist simultaneously in multiple states due to their property of superposition, a feature that gives quantum computers their unique and powerful capabilities.
Quantum gates manipulate qubits much like logic gates manipulate bits in classical computing. Key quantum gates include the Hadamard, Pauli, controlled, and phase gates, which are essential for executing complex quantum algorithms. By combining these gates, quantum circuits are created, linking the software layer (such as compilers) to the hardware (quantum processors). Today, several cloud providers, including IBM Quantum \cite{chow2021ibm}, Amazon Braket \cite{gonzalez2021cloud}, and Microsoft Azure \cite{prateek2023quantum}, provide access to quantum computing resources. To run a quantum algorithm, users submit their designs to a quantum compiler, such as Qiskit \cite{qiskit2024}, which optimizes and schedules the circuits for specific quantum hardware.

Although quantum computers have shown significant advancements, designing quantum circuits remains resource-intensive and requires many iterations to achieve optimal functionality. As a result, quantum circuit designs are highly valuable and are recognized as a critical IP \cite{aboy2022mapping}, facing similar threats as the classical IC designs. 
This paper primarily focuses on the IP threats to quantum circuits during the compilation process.    
The compilation process transforms the original quantum circuit into one compatible and can be executed with the quantum computer. This includes substituting all gates with gates supported by the quantum computing hardware and adding swap gates so that the entanglement among physical qubits supports the multi-qubit gates in the circuit. In addition, various optimizations can be incorporated to improve the performance of compiled circuits. 
However, vulnerabilities exist in this process, as malicious compilers could potentially access and misuse the quantum designs, such as insertion of trojans \cite{das2023trojannet, roy2024hardware} and counterfeiting of quantum designs \cite{yang2024multi}. As these quantum circuits represent valuable IP \cite{aboy2022mapping}, it is essential to protect them from unauthorized use and IP theft.

In IC design, split manufacturing is commonly used to protect against IP theft by ensuring that no single foundry has access to the complete design, thereby safeguarding the design’s IP \cite{rajendran2013split,xie2015security}. Similarly, the ``splitting" can be adapted for quantum circuits. To conceal information from quantum compilers, the circuit can be divided into two or more sub-circuits, which are then compiled separately using different quantum compilers that support the target quantum computer. In this way, none of the compilers will see the entire circuit. This is analogous to split manufacturing of classical IC chips which is a well-studied area of circuit IP protection \cite{perez2020survey}.

In this paper, we propose a novel method to split quantum circuits for secure compilation and our paper has the contribution as follows,
\begin{itemize}
   \item We propose TetrisLock, a novel split compilation method designed for untrusted compilers, aimed at safeguarding the design of the original quantum circuits. This approach eliminates the need for a trusted compiler in the process.
   \item We introduce a process to add random quantum gates that mask the correct function while ensuring the obfuscated circuit retains the same depth as the original, without adding any depth overhead.
   \item We propose to split the obfuscated quantum circuit with an interlocking pattern, drastically increasing the attack complexity of colluding untrusted compilers.
   \item Experimental results on the \textit{RevLib} benchmark set demonstrate that TetrisLock achieves 0\% depth increase, a ~20\% gate count increase (with a total of 1–4 gates inserted), and less than 1\% accuracy loss for the obfuscated circuits.
\end{itemize}

\vspace{-2mm}
\section{Background \& Related Work}
\vspace{-2mm}

\subsection{Split Manufacturing for Electronic Circuits}
\label{ssec:obfuscation}
\vspace{-2mm}
Split manufacturing of a silicon chip separates the fabrication process across multiple foundries. Hence, no foundry has the full picture of the design, and the design IP is protected. This is important because most IC design companies do not own fabs and have to outsource the tape-out process, which puts the IP in the chip designs at risk. Hence, the primary security goal of split manufacturing is to prevent any foundry from inferring the unseen portion of the design from the portion that they are presented with.

There are multiple ways to split chip designs given the manufacturing technology. For ordinary silicon chips, the manufacturing can be split into a front end of line (FEOL) process and a back end of line (BEOL) process, where the FEOL primarily handles transistors and the BEOL handles the metal layers for interconnection. The FEOL can be done in an untrusted fab with advanced technology node. Since the BEOL usually does not require a technology node as advanced as the FEOL, it can be done in a trusted fab, hiding the interconnection among transistors to the untrusted fab thus preserving the IP in the chip design \cite{perez2020survey}. The feasibility of split manufacturing has been demonstrated in multiple research articles \cite{vaidyanathan2014building, vaidyanathan2014efficient, hill2013split}. For example, in heterogeneously integrated 3D chip designs, it is often possible to split them into a “control” layer that does not need very advanced technology nodes and can be manufactured by a trusted foundry, and a “computation” layer that requires advanced technology nodes offered by an untrusted external foundry \cite{valamehr20133}. 


\vspace{-2mm}
\subsection{A Primer on Quantum Computing  } \label{ssec:quantum_circuits}
\vspace{-2mm}
This section provides an overview of the fundamentals of quantum computing, detailing the essential components such as qubits, quantum gates, and quantum circuits. 

\subsubsection{Qubits}
The quantum bit (a.k.a. qubit) is the fundamental building block of quantum computing and is conceptually similar to the classical computing bit. A qubit has two basis states, denoted by the bracket notation as $\ket 0$ and $\ket 1$. 
The basis states for one qubit can be expressed as two-dimensional vectors, e.g., $\ket 0 = [1, 0]^T$ and $\ket 1 = [0, 1]^T$. As a result, the state $\ket \psi$ above can be written as $\ket \psi = \alpha \ket 0 + \beta \ket 1 = [\alpha, \beta]^T$. For multi-qubit states, a similar situation exists. More specifically, there are $2^n$ basis states in the space of $n$-qubit states, ranging from $\ket{0\dots 0}$ to $\ket{1\dots 1}$, and a $n$-qubit state $\ket \psi$ can be expressed by:
\vspace{-3mm}
\begin{equation*}
  \ket \psi = \sum_{i = 0}^{2^n - 1} a_i \ket i
\end{equation*}
where $\sum_{i = 0}^{2^n - 1}|a_i|^2 = 1$.

\subsubsection{Quantum Gates}

Similar to classical logic gates, the fundamental actions at the logic level in quantum computing are performed by quantum gates. These gates carry out unitary operations, which transform the states of input qubits. Quantum algorithms are composed of sequences of these quantum gates, designed to evolve input qubits into desired quantum states. For example, A quantum gate $U$ must satisfy the equations $U U^\dagger = U^\dagger U = I$  ($U^\dagger$ is the conjugate transpose of $U$), meaning that a quantum gate must be a unitary operation. A quantum gate $U$ operating on a qubit $\ket \psi$ can be written down as $\ket \psi \rightarrow U \ket \psi$. 
Single qubit gates operate on a single qubit and are analogous to the elementary logic gates in classical computing, such as the NOT gate. These gates manipulate a qubit by rotating its state vector on the Bloch sphere, a geometric representation of the qubit's state. The most common single qubit gates include the Pauli gates (X, Y, Z), the Hadamard gate (H), and phase shift gates (S, T).

Multi-qubit gates are essential in quantum computing because they enable interactions between qubits, facilitating complex operations in quantum algorithms. They manipulate the states of two or more qubits simultaneously, creating entanglement, a uniquely quantum phenomenon where qubit states become interdependent. For example, the Controlled-NOT (CNOT) gate operates on two qubits: a control qubit and a target qubit. The control qubit would determine whether the operation is applied to the target qubit. The target qubit's state is flipped by the control qubit in the $\ket 1$ state. 
Many quantum algorithms rely heavily on the CNOT gate for their implementation, especially in constructing entangled states and performing conditional operations.




\subsubsection{Quantum Gate Reversibility}
\label{sec:quantum_reverse}
Quantum gate reversibility is a fundamental property of quantum computing, implying that the quantum operations could be canceled out by applying the computation in reverse order. Unlike classical logic gates, which can lose information (e.g., an AND gate reduces two input bits to one output bit), quantum gates are inherently reversible, meaning that their operations can be undone to recover the original state of the system. This reversibility arises from the unitary nature of quantum gates. For example, if a quantum circuit applies a gate $U$ followed by another gate $V$, the reversal of the circuit involves applying $V^\dagger$ followed by $U^\dagger$. This property is critical to the random gate insertion technique that enhances the security for split compilation.

\subsubsection{Quantum Circuits}
A quantum circuit is a sequence of quantum gates arranged to perform a computation on one or more qubits. It typically starts with qubits initialized in a known state (e.g., $\ket{0}$), followed by the application of single and multiple qubit gates. The circuit can be visualized as a timeline with wires representing qubits and gate operations along them.

\subsubsection{Quantum Circuit Compilation}
Quantum circuit compilation transforms a circuit representing a quantum algorithm into a hardware-executable circuit. This process adapts circuits to the specific topology and error characteristics of quantum hardware, similar to compiling classical computer programs but under unique constraints like superposition and entanglement. Compilation ensures efficient and accurate execution of quantum algorithms, bridging the gap between algorithm designs and physical implementation on quantum processors.


\subsection{Related Work and Limitations}

\vspace{-2mm}


Multiple avenues have been explored by researchers to secure quantum circuits against third-party compilers. These include random reversible circuit insertion \cite{das2023randomized, suresh2021short, das2024secure}, locking the quantum circuit by adding extra key qubits \cite{topaloglu2023quantum}, and splitting the quantum circuit and compiling each portion in a different compiler \cite{saki2021split}. In \cite{das2023randomized} and \cite{das2024secure}, a random reversible quantum circuit was generated and inserted at the front, middle, or end of the original quantum circuit to achieve obfuscation. To restore the circuit's functionality after compilation, the reverse of the random circuit was added in the same location. A similar strategy was employed in \cite{suresh2021short}. However, the primary limitation of these approaches is that the topology of the original circuit remains fully exposed, making it possible for an adversary to identify the boundary between the original circuit and the inserted random portion. The technique proposed in \cite{topaloglu2023quantum} adopts an approach akin to classical logic locking, where the circuit's functionality is controlled by multiple key bits. For each key bit, an additional qubit is introduced into the quantum circuit. While this method adds a layer of security, it is impractical for real-world applications due to the high cost of qubits and the limited number of qubits available in current quantum computers.

Compared to random circuit insertion and quantum circuit locking, split compilation has some unique advantages. Since each compiler only sees the part of the circuit that it compiles, other parts of the quantum circuit are completely hidden from the compiler. Saki \textit{et. al.} proposed to split the quantum circuit into two or more cascading sections and compile separately \cite{saki2021split}. To ensure compatibility with quantum computers and resist attacks when the same compiler is used for more than one compilation, they add a swap network between each pair of split-compiled quantum circuits. This approach may still suffer from a brute-force attack when the same compiler compiles two adjacent parts of the circuit. As current quantum computers have a very limited number of qubits, it is still feasible for the adversary to try all permutations of the qubit mapping between the two split circuits.
In this work, we present a novel split compilation technique that uses dummy gates and un-matched qubit splitting to mitigate such brute-force attacks.

\vspace{-3mm}
\section{Threat Models}
\begin{figure}[htb]
\captionsetup{font=small} 
    \centering
    \includegraphics[width=0.5 \textwidth]{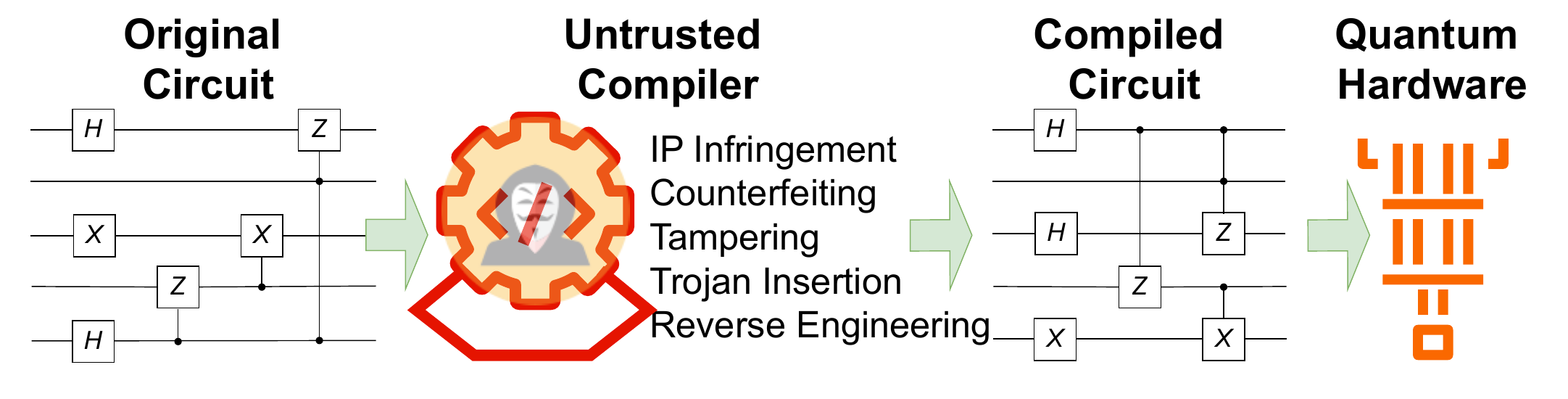}
    \caption{Illustration of the untrusted compiler threat model. During the compile process, the circuit's IP is exposed.}
    \label{fig:threat_model}
\end{figure}

\vspace{-2mm}
In this paper, we consider the third-party quantum compilers as potential adversaries. These third-party quantum compilers offer numerous benefits, such as support for multiple quantum computing platforms, advanced optimization, error correction, etc. \cite{smith2020open,salm2021automating}. Popular third-party quantum compilers include Qulic \cite{smith2020open}, TKET \cite{sivarajah2020t}, etc., and some hardware-specific compilers, like IBM's Qiskit \cite{qiskit2024} and Google's Cirq \cite{hancockcirq}, also support other platforms that can function as a third-party compiler.\footnote{These compilers are only listed as examples to demonstrate the popularity of third-party quantum compilers. We by no means suggest that these compilers are malicious.}
Although third-party compilers offer flexibility, they also pose serious threats to the IP in quantum circuits. During the compilation process, the quantum circuit design is fully exposed to the compiler, leaving its intellectual property vulnerable to infringement or counterfeiting \cite{ghosh2023primer}. An adversary with unauthorized access can replicate the original design and duplicate the circuit’s functionality. These threat models are illustrated in Figure \ref{fig:threat_model}. The threat model considered in this paper is consistent with those used in previous quantum adversaries approaches \cite{das2023randomized,  suresh2021short, das2024secure,topaloglu2023quantum, saki2021split}. 
As far as split compilation is concerned, we also consider the case where the split circuits are compiled using the same compiler, or using colluding compilers.
\vspace{-4mm}
\vspace{-1mm}
\begin{figure*}[htp]
\captionsetup{font=small} 
    \centering
    \includegraphics[width=0.8\textwidth]{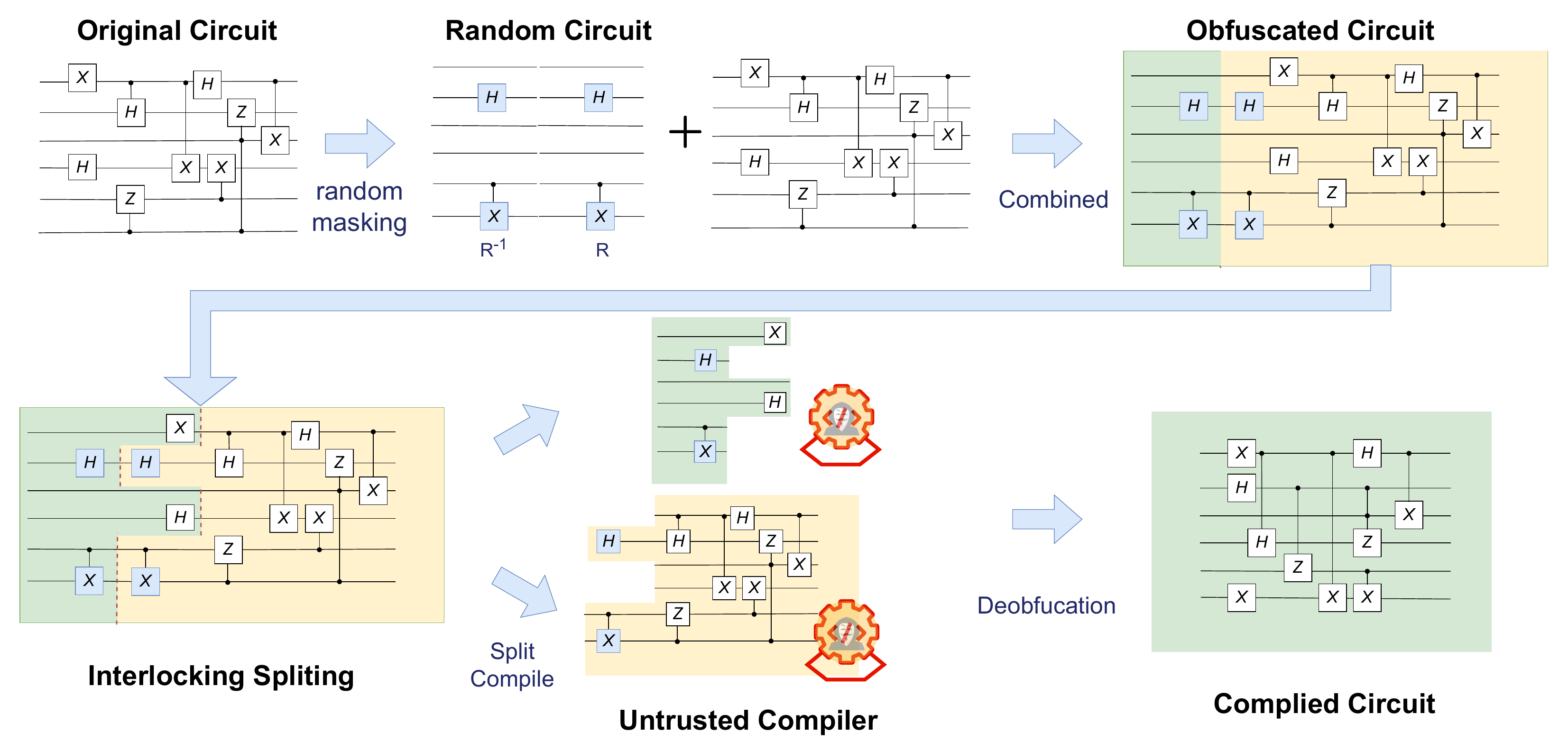}
    \vspace{-1mm}
    \caption{The TetrisLock framework begins by appending the original circuit $C$ with a random circuit $R$, followed by splitting it based on interlocking patterns. Untrusted compiler only has partial of the original circuit. After compilation, the segments are recombined during the de-obfuscation process, fully restoring functionality.}
    \label{fig:defense_flow}
\vspace{-6mm}
\end{figure*}


\vspace{-3mm}
\section{Obfuscation for Quantum Circuits}

Our defense strategy focuses on limiting IP exposure through Quantum circuit obfuscation. We first apply an interlocking method to split the circuit for compilers to process. Since the untrusted compiler does not have the whole piece of the design, the correct functionality of the quantum circuit remains hidden from the compiler. After compilation, a ``de-obfuscation" process is applied to the circuits, restoring the functionality accordingly.  This flow is shown in Figure \ref{fig:defense_flow}.

\subsection{Interlocking Obfuscation}
\label{ssec:qll_method}

TetrisLock, our proposed interlocking obfuscation method, consists of multiple steps to obscure the circuit design before it is sent to the untrusted compiler, as illustrated in Figure \ref{fig:defense_flow}. This process involves segmenting the circuit into interdependent parts, each of which lacks the full functionality of the original circuit before splitting. By obfuscating and distributing these parts across different compilers, we ensure that no single compiler has access to the complete circuit logic, effectively safeguarding the design's integrity and intellectual property.

To facilitate the obfuscation while defending against the collude attacks, we generate a random circuit $R$ to insert before and cover the functionality of the original circuit $C$. To avoid significant increases in circuit depth or gate count, we employ an algorithm to identify empty slots in the original circuits at each layer. The random circuits are then inserted into these empty slots, as detailed in Algorithm \ref{alg:empty_positions}.

\begin{figure}[htb]
    \centering
    \captionsetup{font=small} 
    \includegraphics[width=\linewidth, trim={4cm, 1.2cm, 2cm, 0.5cm}, clip]{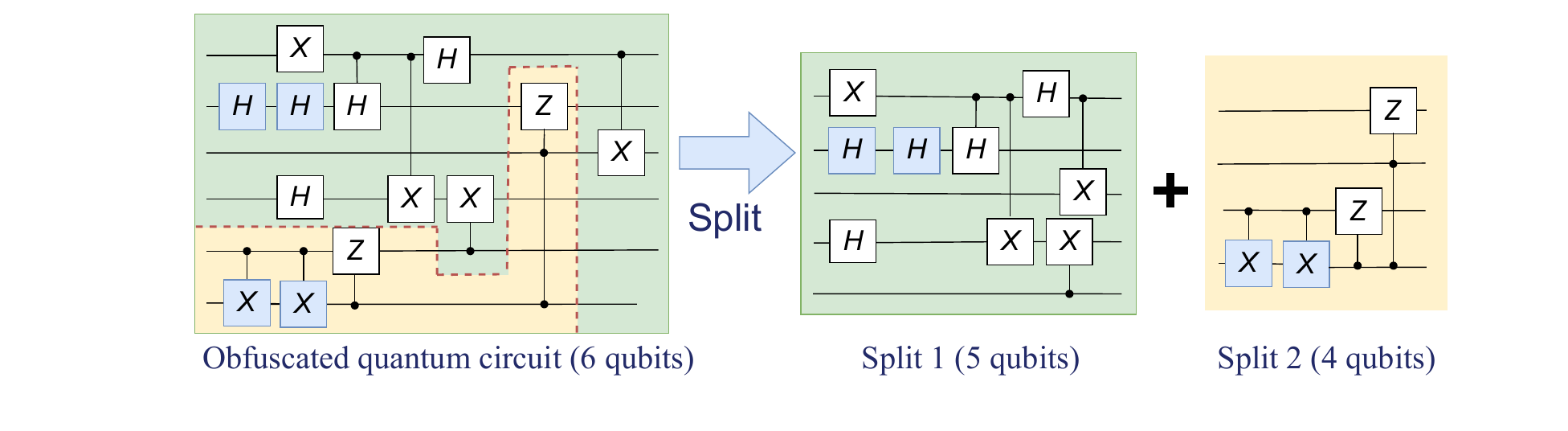}
    \caption{Another way to split the same obfuscated circuit as in Figure \ref{fig:defense_flow} with an interlocking pattern. The two split circuits have different numbers of qubits and not every original qubit needs to be split.}
    \label{fig:inter}
    \vspace{-6mm}
\end{figure}

To reserve the function of the original circuit, we also insert the inverse of the random circuit, denoted by $R^{\dagger}$ or $R^{-1}$. This insertion maintains the original circuit’s functionality due to the reversibility property of quantum circuits discussed in Sec.\ref{sec:quantum_reverse}. Specifically, the reversibility of quantum circuits allows us to implement the inverse by applying the conjugate transpose of each gate in reverse order. For example, if the circuit applies a gate sequence $A$ followed by $B$ as $A \cdot B$, the reversed circuit will apply the gates in the order $B^{-1}$ followed by $A^{-1}$, as $B^{-1}A^{-1}$ where $^{-1}$ denotes the conjugate transpose (adjoint) of each gate matrix.
TetrisLock, which adds $R$ and $R^{-1}$  before the original circuits, will not alter the function but introduce additional gates as masking. It ensures the obfuscated circuit appears substantially different while retaining the original functionality to safeguard the design from unauthorized access during the compilation process.
Thus, the total obfuscated circuit can be represented as $R^{-1}RC$. Ideally, the inverse circuit $R^{-1}$ would cancel out the effect of $R$, ensuring that the modified circuit $C'=R^{-1}RC$ maintains the exact functionality of the original circuit $C$.
\begin{algorithm}[tb]
\DontPrintSemicolon
\SetAlgoLined
\SetNoFillComment
\LinesNotNumbered 
\caption{Random gate insertion into empty positions}\label{alg:empty_positions}
\KwData{\textbf{C}: quantum circuit, \textbf{empty\_pos}: List to store empty positions}
\KwIn{ total\_qubits, gate\_limit}
\tcc{Step1:Get empty positions}
Convert circuit to DAG representation and Extract layers\;

\For{each layer in  \textbf{layers}}{
    Get operations in the current layer\;
    Initialize empty set for used qubits\;
    
    \For{each operation in layer}{
        used\_qubits $\gets$ GetQubitIndices(operation)\;
    }
    empty\_positions $\gets$ sorted(list(all\_qubits $\setminus$ used\_qubits))
}
\tcc{Step2: randomly insertion into circuits}

\ForEach{column $\in$ quantum circuits}{
    \If{added\_gates $>$ gate\_limit}{
    Break\;
    }
    available\_qubits $\leftarrow$ available\_qubits $\cap$ empty\_positions\;
    \If{Random(0,1) $< 0.5$ \textbf{and} find(q1,q2) $==$ true}{
        AddCXGate(circuit, q1, q2) \tcp*[l]{insert CX gate}
        available\_qubits $\leftarrow$ available\_qubits $\setminus \{$q1,q2$\}$\;
        added\_gates $\leftarrow$ added\_gates $+ 1$\;
    }
    \Else{\tcp*[l]{insert X gate}
        random\_pos $\leftarrow$ RandomChoice(available\_qubits)\;
        AddXGate(circuit, random\_pos)\;
        available\_qubits $\leftarrow$ available\_qubits $\setminus\{$random\_pos$\}$\;
        added\_gates $\leftarrow$ added\_gates $+ 1$\;
        }
    }
\KwRet{\textbf{C}: quantum circuit}\;
\end{algorithm}
To defend against the potential attacks to extract the original circuit $C$ from $RC$, we propose an interlocking splitting method between $R$ and $C$. In our approach, portions of the original circuit denoted as $C_l$ are integrated within the random inverse circuit $R^{-1}$, as illustrated in Figure \ref{fig:defense_flow}. Instead of a straight partition, we implement an interlocking boundary to enhance the obfuscation properties of the circuit and thus increase the security level of the process. 
In Figure \ref{fig:inter}, we split the same obfuscated quantum circuit as in Figure \ref{fig:defense_flow} with a different interlocking pattern.
By embedding interlocking gates within the boundary, we create a circuit structure in which, even if an attacker were to isolate the portion of $\tilde{C}_r$, they would still lack access to the complete original circuit $C$. This is because crucial circuit logic segments are now interwoven within $R | C_l$. Consequently, this interlocking strategy significantly reduces the risk of reverse-engineering the original circuit, ensuring the design remains secure.

\vspace{-3mm}
\subsection{De-Obfuscation Process}
\vspace{-2mm}
Our de-obfuscation process combines two separately compiled segments of the circuit as $R^{-1}C_l$ and $RC_r$ together. Since the original circuit was modified by adding a random circuit $R$ along with its inverse $R^{-1}$, the effects of $R$ cancel out, effectively restoring the original functionality of circuit $C$. This combined circuit preserves the original function without revealing it to the separated untrusted compilers.

\vspace{-3mm}
\subsection{Improvements over Prior Work}
\vspace{-1mm}
TetrisLock mitigates significant security vulnerabilities identified in prior work on quantum circuit split compilation \cite{saki2021split} and random circuit insertion for compilation \cite{das2023randomized}.
The primary vulnerability of \cite{saki2021split} is colluding compilers that could collectively reconstruct the original circuit from split segments. Specifically, the adversary can execute an attack by identifying two split circuits with the same number of qubits and subsequently matching the corresponding qubit across both segments. Hence, the complexity of one compiler to find the original circuit is $k_n n!$, where $n$ is the number of qubits, and $k_n$ is the number of segments for the compiled quantum circuits with $n$ qubits from the other compiler.
In contrast, as shown in Figure \ref{fig:inter}, TetrisLock enables the split circuit to have different amounts of qubits. This means that the piece of the circuit from the other compiler can consist of any number of qubits and does not need to be equal to the first piece. The attack complexity is then given by
\begin{equation}
\vspace{-2mm}
\sum_{i=1}^{n_{max}}k_i\sum_{j=0}^{\min{(n,i)}}C_n^jC_i^jj!
\label{eq:complexity}
\end{equation}
let $n_{max}$ denote the maximum number of qubits supported by the target quantum computer, $i$ represent the number of qubits in the other split circuit, and $j$ is the number of qubits that need to be connected. Equation \ref{eq:complexity} calculates the number of possible qubit mapping by considering the enumeration process across each candidate split circuit $k_i$, each possible number of connected qubits $C_n^i$, each combination of connected qubits from both segments $C_n^j$ and each possible mapping between them $j!$. Consequently, the attack complexity of \cite{saki2021split} constitutes only a minor fraction of the total attack complexity in our approach.
Furthermore, the original circuit is structurally obfuscated through the random circuit $R$. This additional layer of masking effectively protects the original design $C$ exposed to the untrusted compiler.

The random circuit insertion approach of \cite{das2023randomized} suggests placing a boundary between $R^{-1}$ and $R$ as $R^{-1} | R C$. This configuration ensured that only the right segment, $RC$, is processed by the untrusted compiler. However, despite the addition of random circuit $R$, an attacker might still be able to identify the boundary between $R$ and $C$. This identification could enable the attacker to recognize and isolate the original circuit $C$ from the one $RC$ sent to the compiler. 
Our approach has the following advantages: (1) No need for a trusted compiler to compile the inserted random circuit. (2) No single compiler have access to the entire original circuit, and (3) Low overhead, as our approach does not increase the overall depth of the quantum circuit.

 


\section{Experiments Evaluation}
\subsection{Experimental Setup}

We conducted our experiments using the IBM Qiskit framework to compile and simulate quantum circuits. We utilized benchmark circuits from the \textit{RevLib} benchmarks \cite{wille2008revlib} for our experiments, which has been widely utilized in prior work on quantum circuit compilation. These benchmark circuits encompass a diverse set of gate operations, with the number of gates ranging from 4 to 32 and qubit sizes varying across 4, 5, 7, 10, and 12 qubits. 
To replicate realistic simulation conditions, we employed the \textit{FakeValencia} backend from Qiskit \cite{qiskit2024}, which incorporates the noise model of the actual \textit{ibmq-valencia} device. All simulations were performed with 1,000 shots to generate statistically significant results. Both the original and the obfuscated circuits were simulated using the same backend, ensuring that any differences observed are attributable to the mechanism rather than variations in the simulation environment.

As described in Section \ref{ssec:qll_method}, we insert a random circuit, composed of a specific set of quantum gates, at the beginning of the original circuit. We strategically selected gate types for insertion based on the operations present in the benchmarks. For instance, in the \textit{RevLib} benchmarks, which predominantly feature arithmetic operations such as adders, ALUs, counters, and comparators, we primarily use NOT (X) and CNOT (CX) gates to obfuscate the circuit and camouflage the circuits from potential attackers. For other types of circuits, such as those implementing Grover's algorithm, we opted to insert Hadamard (H) gates. This tailored approach ensures that the obfuscating mechanism is appropriately aligned with the nature of the operations in each circuit, reducing structural leakage and enhancing the effectiveness of the obfuscation.

\subsection{Metrics for Evaluation}

\textit{Total Variation Distance (TVD)} is a metric used to quantify the distance between two probability distributions. This metric is particularly suitable for quantum circuits measurement because unlike traditional computers, which produce deterministic results, the output of a quantum computing circuit is inherently probabilistic distributions. For example, the output of a 1-bit circuit simulation with noise can be represented as a distribution, such as \{``0'': 95, ``1'': 5\}, based on 100 shots. In this context, TVD measures the discrepancy between the output distributions of the correct (original) circuit and the obfuscated circuit. This discrepancy reflects the effectiveness of the obfuscation in concealing the correct output. It is calculated as the sum of absolute differences between the counts of each outcome in the two distributions, normalized by the total number of shots. The formula for TVD is:
\begin{equation}
    TVD = \frac{\sum_{i=0}^{2^b-1} |y_{i,orig} - y_{i,alter}|}{2N}
\end{equation}
Where $N$ represents the total number of shots in this run, $b$ represents the number of output qubits, resulting in 
$2^b$ possible output types. $y_{i,alter}$ and $y_{i,orig}$ represent the count of value $i$ in the altered and original quantum circuits respectively.


\vspace{-1mm}
\begin{table*}[h]
\captionsetup{font=small} 
\centering
\begin{tabular}{|l|c|c|c|c|c|c|c|c|c|c|}
\hline
\textbf{Circuit} & \textbf{Depth} & \textbf{\CellWithForceBreak{Depth \\ Obfuscated}} & 
\textbf{\CellWithForceBreak{Gate \\ Count}} & \textbf{\CellWithForceBreak{Gate \\ Obfuscated}} & 
\textbf{\CellWithForceBreak{Gate \\ change(\%)}} &
\textbf{Accuracy}  & \textbf{\CellWithForceBreak{Accuracy \\ restored}} & \textbf{\CellWithForceBreak{Accuracy \\ change(\%)}} \\ \hline
mini\_ALU & 8  & 8  & 9 & 11 & 22.2\% & 0.974 & 0.974 & 0.06\% \\ \hline
4mod5    & 5 &5  & 6  & 8 & 33.3\% & 0.973  & 0.967  & 0.6\% \\ \hline
1-bit adder & 5 & 5 & 7  & 8 & 14.2\% & 0.976 & 0.976 & 0.12\% \\ \hline
4gt11    & 13 &13 & 13 & 15 & 15.4\% & 0.986 & 0.983 & 0.30\% \\ \hline
4gt13    & 4  & 4 & 4  & 6.7 & 67.5\% & 0.976 & 0.977 & 0.95\% \\ \hline
rd53     & 16 & 16 & 19 & 22 & 15.7\% & 0.88 & 0.869  & 1.09\% \\ \hline
rd73     & 13 & 13 & 23 & 26 & 13.0\% & 0.892 & 0.884 & 0.73\% \\ \hline
rd84     & 15 & 15 & 32 & 36 &  12.5\% & 0.867 & 0.863 & 0.42\% \\ \hline
\end{tabular}
\caption{Comparison of circuit parameters: depth, count, accuracy, and fidelity change before and after alterations, data shown here are the averages of 20 iterations. The original circuit is from the \textit{RevLib}.}
 \vspace{-6mm}
\label{tab:circuit_parameters}
\end{table*}

\vspace{-1mm}
\subsection{Result Analysis}
\begin{figure}[!htp]
    \centering
    \captionsetup{font=small} 
    \includegraphics[trim={0 2mm 0 0},clip,width=0.44\textwidth]{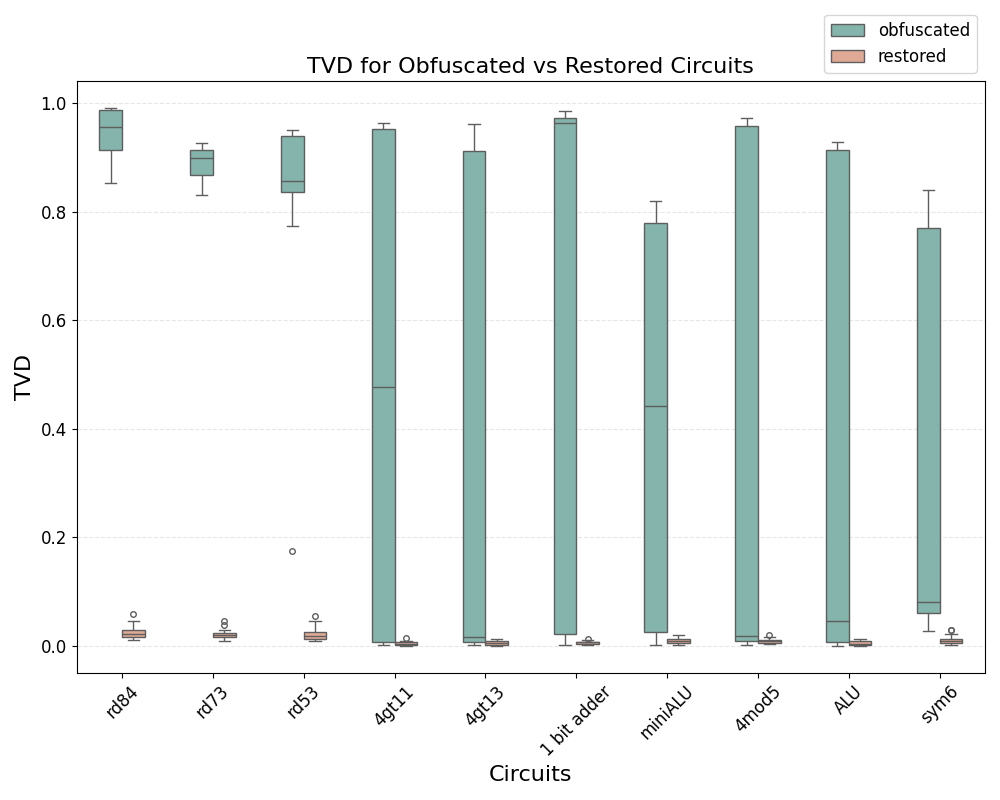}
    \caption{Distribution of Total Variation Distance (TVD) of benchmark circuits: TVD of obfuscated circuit and restored circuit are calculated and shown respectively. Selected circuits are simulated using Qiskit and FakeValencia backend, incorporating noise into the simulation}
    \label{fig:result_vd}
\end{figure}

\vspace{-1mm}
In this section, we present our experimental results, starting with the outcomes from the \textit{RevLib} benchmarks simulated using the Qiskit backend. These results encompass both 1-bit and multi-bit circuit scenarios.
Figure \ref{fig:result_vd} illustrates a comparison of TVD values across different circuits for the obfuscated case and restored case.  TVD is calculated as the variation distance with the theoretical output. For instance, in the case of a 1-bit Adder with 100 shots, we use the outcome as  \{``0'': 100, ``1'': 0  \} as the reference to compare relative distance. 

\vspace{-5mm}
\vspace{-2mm}
First, significant variations in TVD values are observed for the obfuscated circuits, indicating functional alterations caused by the insertion of the random circuit. For circuits such as rd84, rd73, and rd53, the TVD values approach 1, indicating significant changes in the output distribution. This occurs because these circuits produce multi-bit outputs and are relatively large and deep, which provide ample opportunities for the insertion of random gates. More insertion of random gates results in more flips in the output. In contrast, smaller circuits with 1-bit output, such as the 1-bit adder (4-qubit) and 4gt13 (4-qubit) have less significant changes in TVD value. This is because their simpler structure provides limited space for inserting random gates compared to more complex circuits. For the restored case, $R^{-1}$ is applied to the obfuscated circuit $RC$, effectively canceling the disturbance introduced by random circuit $R$. TVD values in this case are consistently close to zero, confirming that the circuits retain the same functionality as the original.
Overall, Figure \ref{fig:result_vd} shows that the obfuscated circuits differ significantly from the original circuits in TVD values. The consistent TVD values across benchmarks indicate that our TerisLock method uniformly impacts all circuits.

\vspace{-1mm}

\subsection{Cost and Overhead Analysis}
\vspace{-1mm}

\subsubsection{Gate Counts and Circuit Depth}

Results regarding gate counts and circuit depth across various circuits are presented in Table \ref{tab:circuit_parameters}, revealing several trends in how alterations impact circuit complexity and performance. First, we applied a random insertion algorithm to ensure that the inserted gates occupy existing empty slots of the circuit layers, keeping the circuit depth unchanged after obfuscation. This demonstrates the effectiveness of the obfuscation method in preserving depth while covering the circuits' functions. 
The number of gates inserted across the circuit benchmarks ranges from 2 to 4, resulting in an average 20\% increase in the total gate count. However, for larger and deeper circuits, the relative impact of these additional gates is minimal, as they represent a small proportion of the original gate count. 

\subsubsection{Fidelity of Unlocked Circuits}
In this section, we evaluate the performance of circuits after their restoration following the split compilation process. Splitting a circuit into two parts and introducing additional gates and layers can affect its accuracy. To assess this, we measured the accuracy (the ratio of correct outcomes to the total number of shots) both before and after the circuits were restored. As shown in Table \ref{tab:circuit_parameters}, most circuits exhibited only a slight decrease in accuracy post-restoration, with fidelity changes typically remaining below 1\%. Overall, while these modifications increase the circuit's complexity, their impact on performance is negligible. This demonstrates that the circuits maintain their robustness and reliability even after undergoing the introduced alterations.

\vspace{-1mm}
\section{Conclusion}
\vspace{-1mm}
In this paper, we introduce TetrisLock, a novel split compilation technique that enhances quantum circuit security by inserting random quantum gates and splitting the circuit using an interlocking pattern. We demonstrate that this approach has a significant advantage over existing split compilation and random gate insertion-based methods. Our experiments across various quantum circuits show that TetrisLock achieves a high functional corruption when obfuscated and has minimal impact on accuracy when the split circuits are correctly recombined. Hence, TetrisLock is a viable solution for protecting quantum circuit IP through split compilation.

\balance

\newpage

\bibliographystyle{IEEEtran.bst}
\bibliography{qref}
\end{document}